\documentstyle{l-aa}
\input{epsf}

\def\beq{\begin{equation}}
\def\eeq{\end{equation}}
\def\bea{\begin{eqnarray}}

\def\eea{\end{eqnarray}}
\def\ds{\displaystyle}

\def\ni{\noindent}

\def\req#1{(\ref{#1})}

\def\ln{\mbox{ln}}

\def\wh#1{\widehat{#1}}

\def\text#1{\mbox{\scriptsize #1}}

\def\barc{\begin{array}{c}}
\def\ear{\end{array}}

\def\bit{\begin{itemize}}
\def\eit{\end{itemize}}

\def\Kbs{{\,{\mbox{~Kb}} {\,s^{-1}\,}}}
\def\eea{\end{eqnarray}}
\def\ds{\displaystyle}

\def\ni{\noindent}

\def\ln{\mbox{ln}}

\def\barc{\begin{array}{c}}
\def\ear{\end{array}}

\def\bit{\begin{itemize}}
\def\eit{\end{itemize}}

\def\and  {\it {et al.} \rm}

\def\etal{{\rm et~al. }}
\def\spose#1{\hbox to 0pt{#1\hss}}
\def\simlt{\mathrel{\spose{\lower 3pt\hbox{$\mathchar"218$}}
     \raise 2.0pt\hbox{$\mathchar"13C$}}}
\def\simgt{\mathrel{\spose{\lower 3pt\hbox{$\mathchar"218$}}
     \raise 2.0pt\hbox{$\mathchar"13E$}}}
\def\beq{\begin{equation}}
\def\eeq{\end{equation}}
\def\bce{\begin{center}}
\def\ece{\end{center}}
\def\bea{\begin{eqnarray}}
\def\eea{\end{eqnarray}}
\def\ben{\begin{enumerate}}
\def\een{\end{enumerate}}

\def\ni{\noindent}

\def\brr{\begin{array}}
\def\err{\end{array}}

\def\calR{{\cal R }}
\def\calD{{\cal D }}
\def\calT{{\cal T }}

\begin{document}

\thesaurus{12.03.1, 12.03.3, 03.13.6, 03.13.2, 03.20.4}

\title{\ni Data Processing and Compression of Cosmic Microwave 
Background Anisotropies on Board the PLANCK  Satellite}


\author{\ni E. Gazta\~{n}aga,   A. Romeo,  J. Barriga,
E. Elizalde\thanks{{\it 
Presently on leave at: }
Department of Mathematics, Room 2-363A, Massachusetts Institute of
Technology, 77 Massachusetts Av., Cambridge, MA 02139-4307, USA.}
}


\institute
{Consejo Superior de Investigaciones Cient\'{\i}ficas (CSIC), 
~Institut d'Estudis Espacials de Catalunya (IEEC),  \\ 
\indent Edf. Nexus-201 - c/ Gran Capit\`a 2-4, 08034 Barcelona,
Spain}

\maketitle

\markboth{Data Processing and Compression of CMB 
Anisotropies on Board the PLANCK  Satellite}{E. Gazta\~naga}


\begin{abstract}

We present a simple way of coding and compressing 
the data on board the Planck
instruments (HFI and LFI)
to address the problem of the on board data reduction. This is
a critical issue in the Planck mission. The total 
information that can be downloaded to Earth is 
severely limited by the
telemetry allocation. This limitation could reduce the amount of 
diagnostics sent on the stability of the radiometers
and, as a consequence, curb the final sensitivity of the CMB 
anisotropy maps.
Our proposal to address this problem 
consists in taking differences of consecutive circles
at a given sky pointing. To a good approximation,
these differences are 
independent of the external signal, 
and are dominated by thermal (white) instrumental noise. 
Using simulations and analytical predictions we show that
high compression rates, $c_r \simeq 10$, can be obtained with minor
or zero loss of CMB sensitivity. Possible effects of
digital distortion are also analized.
The proposed scheme  allows for flexibility to optimize
the relation with other critical aspects of the mission.
Thus, this study constitutes an important step towards 
a more realistic modeling of 
the final sensitivity of the CMB temperature anisotropy maps.


\end{abstract}


\section{Introduction}

The PLANCK Satellite is designed to measure temperature fluctuations
in the Cosmic Microwave Background (CMB) with a precision of
$\simeq 2 \mu K$, and  angular resolution of about 5 arcminutes.
The payload consists of a 1.5-2.0 m Gregorian telescope which feeds
two instruments:
the High Frequency Instrument (HFI) with 56
bolometer arrays operated at $0.1K$ and frequencies of $100-850$GHz
and the Low Frequency Instrument (LFI) with 
56 tuned radio receivers arrays
operated at $20K$ ($4K$) and frequencies of $30-100$ GHz
(see  http://astro.estec.esa.nl/SA-general/Projects/Planck/ for
more information).

Data on board PLANCK
consist of $N$ differential temperature measurements, spanning a range
of values we shall call $\calR$. Preliminary studies and telemetry allocation 
indicate the need for compressing these data
by a ratio of  $c_r \simgt 10$. Here we will consider
under what conditions it might be possible to achieve such
a large compression factor.

A discretized data set can be represented by a number of bits,
$n_{\text{bits}}$,
which for linear Analogue-to-digital converters (ADC)
is typically given by the maximum range $N_{max}$:
$n_{\text{bits}}=\log_2{N_{max}}$.
If we express the joint probability for a set of N measurements
as $p_{i_1,\dots, i_N}$,
we have that the Shannon entropy per 
component of the data set is:

\beq
h \equiv - {1\over{N}} \sum_{i_1,\dots, i_N} p_{i_1,\dots, i_N}
\log_2(p_{i_1,\dots, i_N}).
\label{h}
\eeq
Shannon's theorem states that $h$ is a lower bound to the average length of
the code units.
We will therefore  define the  theoretical (optimal) compression rate as
\beq
c_{r,opt} \equiv {n_{\text{bits}}\over{h}}
\label{cr}
\eeq
For a uniform
distribution of $N$ measurements we have $p_i=1/N$ and $h= log_2 N$,
which equals the number of bits per data. Thus: it is not possible to
compress a (uniformly) random distribution of  measurements.

Gazta\~naga \etal (1998), 
have argued that a {\it well calibrated signal} 
will be dominated by thermal (white) noise
in the instrument: $\sigma_e \simeq \sigma_T$ and therefore
suggested that the digital resolution
$\Delta$ only needs to be as small as the instrumental RMS white noise:  
$\Delta \simeq \sigma_T \simeq 2 mK$. The nominal $\mu K$ pixel 
sensitivity will only be achieved after averaging (on Earth).
This yields compression rates of $c_{r,opt} \simeq 8$. 
On the other hand  Maris \etal (1999) have used the same
formalism as Gaztanaga \etal but fixing 
the final dynamical range $\calR$ to some fiducial values, 
so that the digital resolution is then given by 
$\ds\Delta = {\calR\over{2^{n_{\text{bits}}}}}$, independently
of $\sigma_T$. Again, assuming a well calibrated signal 
dominated by thermal (white) noise,
this approach yields smaller compression rates of $c_{r,opt} \simeq 4$,
as it is obvious from the fact that the digital 
resolution  is larger ($\Delta$ is smaller).
In both cases, the effect of the CMB signal (eg dipole) and other
sources (such as the galaxy) have been ignored.

Several questions arise from these studies. What is the optimal
value of $\Delta$ and what are the penalties (distortions) involved
when using large values of $\Delta$?  Moreover, can the data 
gathered by the on board instruments be really modeled as
a white noise signal? In other words, are the departures 
from Gaussianity (due to the galactic, 
foregrounds, dipole and CMB signals) important? This latter question is
closely related  to the way data will be processed (and calibrated) 
on board, for example:  
if and how the dipole is going to be used for calibration.
These issues together with the final instrument specifications
seem to play an important role on the final range of values $\calR$ and, 
therefore, the possible compression rates. 
This is somehow unfortunate as compression
would then be related in a rather complicated way to the nature of the
external signal and also to critical issues of
the internal data processing issues.

Here we shall present a simple way of coding the on board data that will  
solve the lossless compression problem in a much simpler way. This
will be done 
independently of the internal calibration or the nature of the external
signal (CMB or otherwise). We will
also address the issue of the digital distortion introduced (the penalty)
as a function of the final compression (the prize).

In section \S 2 we give a summary of some critical issues 
related to the on-board data. Our coding and
compression proposals are presented in \S 3, while simulations
are dicussed in \S 4. We end up with some concluding remarks.


\section{ON BOARD DATA}

\subsection{Data Rate, Telemetry and Compression} 

To illustrate the nature of the compression  problem
we first give some numbers related to the LFI.
Similar estimations apply to the HFI.

\begin{table}

\begin{center}

\begin{tabular}{llllll}
\scriptsize {GHz} & \scriptsize{FWHM} & $\sigma_T$(mK) & T (mK) & Det. & $\Kbs$\\ \hline \hline
30 & 33' & 2.8 & -30-61  & 4 & $9.3$ \\ \hline 
44 & 23' & 3.2 & -30-138  & 6 & $13.9$  \\ \hline
70 & 14' & 4.1 & -20-340  & 12 & $27.8$\\ \hline
100 & 10' & 5.1 & -10-667  & 34 & $78.8$\\ \hline
\hline
\scriptsize {TOTAL} &  &  & -30-667 & 56 & $130$ \\ \hline \hline
\scriptsize {+LOAD} &  &  &   & 112 & $260$ \\ \hline 
\end{tabular}
\end{center}
\caption{
Parameters for the radiometers: a) central frequency $\nu$
(bandwidth is 20\%); b) angular resolution (beam FWHM); c) RMS thermal
noise expected at 6.9 ms (144.9 Hz) sampling; d) range of temperatures 
expected from the sky (Jupiter, dipole, S-Z); e) number of detectors
(2x horns); f) total data rate at 6.9 ms (2.5 arcmin).}
\label{radio}
\end{table}

According to the PLANCK LFI Scientific and Technical Plan 
(Part I, \S 6.3, Mandolesi \etal 1998)
the raw data rate of the LFI is $r_d \simeq 260 \Kbs$. This assumes: 
i) a sample frequency of 6.9 ms or $f_{sampl}=144.9$ Hz, 
which corresponds to 2.5 arcmin in the sky,
1/4 of the FWHM at 100 GHz,
ii) $N_{detec}=112$ detectors: sky and reference load temperature for 56
radiometers. 
iii) $n_{\text{bits}}=16$ bits data representation.
Thus that the raw data rate is:
\beq
r_d = f_{sampl} \times  N_{detec} \times n_{\text{bits}} \simeq 259.7 \Kbs.
\eeq
The values for each channel are shown in Table 1. 
A factor of two reduction can be obtained by only transmitting the 
difference between sky and reference temperature. 
To allow for the recovery of 
diagnostic information on the separate stability of the 
amplifiers and loads,
the full sky and reference channels of a single radiometer
could be sent at a time, changing the selected radiometer from time to time
to cover all channels (Mandolesi \etal 1998). 

Note that the sampling resolution of 6.9 ms corresponds to 2.5
arcmin in the sky, which is smaller than the
nominal FWHM resolution. 
Adjacent pixels in a circle could be averaged on-board to obtain the nominal 
resolution (along the circle direction). 
In this case the pixel size should
still 
be at least $\simeq 2.5$ smaller that the FWHM to allow for a proper
 map reconstruction. 
Note that each circle in the sky will be separated 
by about 2.5' so even after this averaging along the circle scan
 there is still a lot of redundancy 
across circles. For
pixels of size $\theta \simeq FWHM/2.5$ along the circle scan
the total scientific rate could be reduced to $r\simeq 67 \Kbs$
 (or $134 \Kbs$ with some subset information of the ref. load).

 The telemetry allocation for the
LFI scientific data is expected to be $r_t=20 \Kbs$.
Thus the target compression rates
are about:

\beq
c_r = {r_d\over{r_t}} \simeq 3- 13,
\eeq
depending on the actual on-board processing and
requirements. 

\subsection{Scanning and Data Structure}

The Planck satellite spins with a frequency $f_{spin}=1$ rpm
so that the telescope (pointing at approximately right angles 
to the spin axis)
sweeps out great circles in the sky. Each circle is scanned at the
same position in the sky $\theta$ for 
over 2 hours, so that there are 
120 images of the same pixel (the final number might
be different but this is not relevant here). 
We can write the whole data in each pointing $\theta$ as a matrix:
\beq
D_{k,\alpha}(\theta) = S_{k,\alpha}(\theta) + \eta_{k,\alpha}
\label{Dka}
\eeq
where $S$ stands for the external {\it signal} (CMB, galaxy, foregrounds)
and $\eta$ stands for the internal (eg, instrumental)
noise. The $k$ index labels the number of spins
in that pointing and $\alpha$ labels the positions within the circle.
Each measurement is mostly 
dominated by instrumental noise, 
$\sigma_T \sim 2 mK$ (see Table 1)
rather than by the CMB noise ($\sigma_{CMB} \simeq 10^{-2} mK$).
If this noise (at frequencies smaller
than $f_{spin}$) is mostly 
thermal,  one could then say that there is no need for compression, 
as we can just average those 
120 images of a given pixel in the sky
and only send the mean downwards to Earth. 
The problem is that 
one expects $1/f$ instabilities to dominate the
instrument noise at frequencies smaller than $\sim 0.1$ Hz. 
Compression is only required when we want to keep these 120 images
in order to correct for the instrument instability in the data 
reduction process (on Earth). 

\subsection{Dynamic Range \& Sensitivity}
\label{sensitivity}

The rms standard deviation level in the CMB anisotropies
is expected to be  of a few tens of $\mu K$. These
anisotropies will be mapped with a $\simeq 1 \mu K$ 
resolution.
But the final dynamic range for the measured temperature differences
per angular resolution pixel will be $\Delta T \simeq 1 \mu K- 1K$. The 
maximum resolution
of $\simeq 1 \mu K$ will only be obtained after averaging all data.
The highest value $\simeq 1K$ is given by the hottest source that we 
want to keep (not saturated) at any of the frequencies.
Positive signals from Jupiter, which will be used for calibration,
can be as large as $\simeq 0.7K$ at 100 GHz. Other point sources and
the Galaxy give intermediate positive values. 
Negative differences (with respect to the 
mean CMB $T \simeq 2.7K$), of the order of a few $mK$, can be originated 
by the dipole, the relative  velocity  between the
satellite velocity and the CMB rest frame. The Sunyaev-Zeldovich effect
can also give a negative signal of few $10 mK$. Thus the overall range of
external temperature differences could be $-30 mK$ to $1 K$.
The internal reference load will also be subject to variations
which have to be characterized. The instrument dynamic range
will depend on the final design of the radiometers and its
internal calibration. This is not well understood yet and
it is therefore difficult to assess how it will affect
the on-board information content.

Planck LFI radiometers are modified Blum correlation  
receivers (see Blum 1959). 
Both LFI and HFI radiometers have an ideal white noise 
sensitivity of

\beq
\calT =\frac{\sigma_\nu}{\sqrt{\nu~ \tau}} = \frac{\sigma_T}{\sqrt{N}}
\label{deltaTCMB}
\eeq
where $\tau$ is the integration time, $\nu $ is the band width
(about $20\%$ of the central frequency 
of the channel for the LFI) and $\sigma_T$ is a characteristic
 rms noise temperature. 
The values of $\sigma_T$ (shown in Table 1) correspond 
to the equivalent noise in a sampling interval, and $N$ 
above is the number of such samplings (or pixels) at a given sky position.  
The final target
sensitivity required by the 
Planck mission to ``answer'' many of our
cosmological questions about the CMB is about 
$\calT_{CMB} \simeq 10^{-6} K$. Thus, 
we need to integrate over about $N \simeq 10^6$ elements 
(i.e. pixels) with the thermal noise shown in Table 1.
This, of course, is just an order of magnitude estimation as
the detailed calculation requires a careful consideration of
the removal of instrument instabilities and the
use of multiband frequency to subtract the different contaminants.

As pointed out by Herreros \etal (1997)
the temperature digital resolution
should be given by the receiver noise $\sigma_T$ on the sampling
time 6.9 ms (or corresponding value if there is some on-board averaging)
and not by the final target sensitivity.
At the end of the mission, each FWHM pixel will have been measured 
many ($\simeq 10^6$) times. Thus a higher resolution 
of $\Delta T \simeq 1 \mu K$
is not necessary on board, given that the raw signal is dominated 
by the white noise component. This higher resolution will be later obtained
by the pixel averaging (data reduction on Earth). 
Using an unnecessary 
high on-board temperature resolution (eg a small $\Delta$) will 
result in a larger Shannon entropy (eg $h \propto log(1/\Delta)$)
which will limit even more the amount of scientific and
diagnostic information that can be download to Earth.

\subsection{Instrumental Noise \& Calibration}

We can distinguish two basic components for the receiver noise: the white
or thermal noise, and the instabilities or calibration gains
(like the $1/f$ noise). An example is given by 
the following power spectrum of 
frequencies $f$:

\beq
P(f) = A \left(1 + {f_{knee}\over{|f|}}\right).
\label{pk}
\eeq
The 'knee' frequency, $f_{knee}$, is expected to be $f_{knee} \simeq 0.005$ Hz
for a 4K load or  $f_{knee} \simeq 0.06$ Hz
for a 20K load.
The expected RMS thermal noise, $\sigma_T \propto A$
at the sampling frequency (2.5 armin), 
is listed in Table 1. The lowest value is
given by the 30 GHz channel and could be further reduced to $\simeq 1 mK$
if the data is averaged to FWHM/2.5 to obtain the nominal resolution.
The larger values in the dynamical range can be affected by the calibration 
gains. This is important and should be carefully taken into account if
a non-linear ADC is used, as gains could then
 change the relative significance
of measurements (eg, less significant bits shifting because of gains).
In fact, a $1/f$ power spectrum integrated from the knee-frequency 
($f_{knee}$) for a time $T$, 
gives a rms noise that diverges with $T$. 
The integration (or sampling) over a single pixel could
be modeled as some (effective) sharp-k window of size $f_{max}$:

\beq
\sigma_{1\/f}^2= {\sigma_T^2\over{f_{max}}} \int_{1/T}^{f_{max}} 
df \, {f_{knee}\over{f}} 
= \sigma_T^2 {f_{knee}\over{f_{max}}} \, \ln{(T \, f_{max}})
\label{sigma1f}
\eeq

For a $T \simeq 1$ year mission the contribution from the $1/f$ noise
in pixels averaged after
succesive pointings $f_{max} \simeq 10^{-4}$ and we have
 $\sigma_{1\/f}^2 \simeq 10^4 \sigma_T^2$! This illustrates
why the calibration problem is so important and makes a large
dynamic range desirable. Averaging pixels at the
spin rate, $f_{max} \simeq f_{spin}$, gives 
$\sigma_{1\/f}^2 \simeq 10 \sigma_T^2$, this is not
too bad for the dynamic range, but it corresponds to a mean value and
there could be more important instantaneous or temporal gains.
Drifts with periods longer than the spin period 
(1 rpm) can be removed by requiring 
that the average signal over each rotation at 
the same pointing remains constant.
Drifts between pointings (after 2 hours) could be reduced by using the 
overlapping pixels. 
All this could be easily done on-board, while a more careful
matching is still possible (and necessary) on Earth. 
This allows the on-board gain
to be calibrated on timescale larger than 1 min with an accuracy given by 
$\sigma_T$. Additional and more carefull in-flight
calibration can also be done using the the signal from external planets
and the CMB dipole. Although this is an interesting possibility 
for the on-board reudction we will present below a simpler 
and more efficient alternative.

\section{CODING \& COMPRESSION}

We will assume from now on that the external signal does not vary 
significantly with time during a spin period (1 minute), i.e.
$S_{k,\alpha}(\theta) \simeq S_{k+1,\alpha}(\theta)$, so that
Eq.[\ref{Dka}] yields:
\beq
D_{k+1,\alpha}(\theta) \simeq S_{k,\alpha}(\theta) + \eta_{k+1,\alpha}.
\eeq
Consider now the differences $\delta$ between the circle scans 
in two consecutive spins of the satellite:
\beq
\delta_{k,\alpha}(\theta) \equiv D_{k+1,\alpha}(\theta)-D_{k,\alpha}(\theta)
\simeq \eta_{k+1,\alpha} - \eta_{k,\alpha}.
\label{delta}
\eeq
These differences are independent of the signal $S_{\alpha}(\theta)$
and are just given by a combination of the noise $\eta$.
Obviously the above operation does not involve any information
loss as the set of  original data images $(D_{k,\alpha} ~,~ k=1,120)$
can be recovered from one of the full circles,
say $D_{1,\alpha}$, and the rest of the differences 
$(\delta_{k,\alpha} ~,~ k=2,120)$. Occasionally,
the external signal could vary significantly during 1 minute
(eg cosmic rays, a variable star or some outbursts). This will
not result in any loss of information but will change
the statistics (and therefore compressibility) of the 
the differences $(\delta_{k,\alpha}$. Here we assume that
the overall statistics are dominated by the instrumental noise.
A more detailed study will be presented elsewhere.

What we propose here is to compress the above noise differences
$\delta_{k,\alpha}(\theta)$ before downloading them to Earth.
This has several advantages over the
direct compression of the $D_{k,\alpha}$:

\begin{itemize}
\item $\delta_{k,\alpha}$ are independent of the input signals, which 
are in general non-Gaussian, eg galaxy, foregrounds, planets...
\item The new quantity to be compressed should approach a (mutivariate)
Gaussian, as it is just instrumental noise.
\item this scheme is independent of any on board calibration
or processing.
\item $\delta_{k,\alpha}$ should be fairly homogeneous (the radiometers
are supposed to be fairly stable over time scales of 1 minute), so that
compression rates should be quite uniform .
\item because of the reasons above there is a lot of
flexibility on data size and processing requirements. 
For the raw data estimated in Table 1 of $c_d \simeq 260 \Kbs$
it will take about $\simeq$ 2 Mbytes to store a full revolution. Thus, 
compression of a few circles at a time might be possible with a
$\simeq$ 16 Mbytes on-board RAM memory. 
\item The resulting processing will be signal lossless 
even if the noise is binned with a low resolution before compression.
This is not clear when $D_{k,\alpha}$ are used instead.
\end{itemize}

In the last point, digital binning of the noise $\delta_{k,\alpha}$
could affect the final 
sensitivity of the  mission by introducing additional 
digital distortion or discretization
noise, which could add to the instrumental noise in a significant way.
We will later quantify this.

We will further assume that the noise  $\eta_{k,\alpha}$
in Eq.[\ref{delta}] is not a function of the position in the sky
but just a function of time. Thus we will assume that $\eta_{k,\alpha}$ are
a realization of an stochastic (multivariate) Gaussian process with
a given power spectum: $P(f)$, eg Eq.[\ref{pk}]. We then have:
\beq
\delta_{k,\alpha} = \eta_{k+1,\alpha} - \eta_{k,\alpha},
\label{delta2}
\eeq
so that  $\delta_{k,\alpha}$ will also be Gaussian, but with a different 
power spectrum. 
To a good approximation the noise $\delta $ will be almost white or thermal,
as differences between components of adjacent 
vectors (circles) are separated by 1 min. ($f_{spin}\approx 0.02 Hz.$) 
which is comparable to or larger than the 
typical  $f_{knee}$ frequencies ($f_{knee} \simeq 0.005$Hz for 4K load
in the LFI).
From now on we will assume, for the sake of simplicity, that $\delta$ is
a purely white noise with $\sigma_\delta \simeq \sqrt{2} \sigma_\eta$ 
($\simeq 3 mK$). Deviations from this assumption are studied in 
Appendix A. 

To estimate the  entropy associated with $\delta_{k,\alpha}$ and its
corresponding $c_{r,opt}$ in Eq.[\ref{cr}] we need to know
how $\delta$ is discretized, i.e., what is $\Delta$ in Eq.[\ref{crdelta}].
This value will in principle be given by the ADC hardware: $\Delta_{ADC}$.
The details of 
the ADC in each instrument will be driven by the electronics, 
the final target of
temperature range $\calR$ and the internal calibration processes.

\subsection{Digital Distortion}

In order to make quantitavive predictions we need to know the
ADC details, i.e., how the on-board signal will be digitalized.
To start with, we will take the digital resolution
$\Delta$ to be a variable. The
noise differences could be subject to a further (on board)
digitalization  (in general with the possibility of
$\Delta \gg \Delta_{ADC}$). This would allow
the compression target to be independent of other mission critical points. 
If the ADC digital resolution
is significantly larger  than the value of
$\Delta$ under consideration ($\Delta_{ADC}<\Delta$) 
the binned data will suffer
an additional {\it digital distortion}, which will add to the 
standard ADC distortion (which will be probably given by other
instrumental considerations). 
In general we represent the overall {\it digital distortion}
by ${\cal D}$, which is defined as
\beq
\calD ~\equiv~
{ D_{\text{err}}^2 \over \sigma_{\delta}^2 } \equiv  
{ \langle (\wh{\delta}-\delta~)^2 \rangle \over{\sigma^2_\delta}},
\label{defD2}
\eeq
where $\wh{\delta}$ is the discretized version of $\delta$ and 
$\langle \dots \rangle$
is the mean over a given realization. 
It is well known (see e.g. \S 5 in Gersho \& Gray 1992)
that in the limit of small $\lambda\equiv \Delta/\sigma$,
the {\it digital distortion} of a signal is simply given by 
\beq
\calD ~\equiv~
{ D_{\text{err}} \over \sigma^2_{\delta} }
 \simeq { \Delta^2\over 12\sigma^2_{\delta} }
={\lambda^2 \over 12},
\label{D2smallla}
\eeq
i.e., ${\cal D}$ is proportional to the digital resolution 
in units of the rms (white noise) deviation. The rms $\sigma$ of the 
discretized version of $\delta$, which we shall call $\wh{\delta}$, is 
\beq
\begin{array}{lll}
\sigma_{\wh{\delta}}&=&\ds
\sqrt{ 1+ { D_{\text{err}}^2 \over \sigma_{\delta}^2 }
+2 { \langle \epsilon \delta \rangle \over\sigma_{\delta}^2 } }  \,
\sigma_{\delta} \\
&\simeq& \ds \sqrt{1+{ D_{\text{err}}^2 \over \sigma_{\delta}^2 } } 
~ \sigma_\delta 
\simeq \sqrt{ 1+{\lambda^2\over 12} } ~\sigma_\delta,
\end{array}
\eeq
where $\epsilon\equiv \wh{\delta}-\delta$ and
$\langle \epsilon \delta \rangle$ denotes the correlation between 
this quantity and $\delta$, which is usually small. 
The discretized field has a larger rms
deviation than the original one.

As mentioned in \S\ref{sensitivity}, the final signal sensitivity,
$\calT_{CMB}$ of the survey
will only be achieved on Earth after averaging many
observations, destripping, galaxy and foreground removal, etc. 
Eq.\req{deltaTCMB} shows that 
this sensitivity should be proportional to a combination
of the thermal noises of each
instrument ---$\sigma_T$--- and, therefore, to $\sigma_\delta$.
Thus, the relative effect of the discretization 
on the mission sensitivity is just given by the ratio
\beq
\begin{array}{lll}
\ds{\Delta \calT_{CMB}\over{ \calT_{CMB}}} = 
{ \sigma_{\wh{\delta}} - \sigma_{\delta} \over \sigma_{\delta} }&=&\ds
\sqrt{ 1+ { D_{\text{err}}^2 \over \sigma_{\delta}^2 } 
+2 { \langle \epsilon \delta \rangle \over\sigma_{\delta}^2 }  } 
-1 \\
&\simeq&\ds \sqrt{1+ { \lambda^2\over 12} } -1.
\end{array}  
\label{senCBM}
\eeq
The approximate form is valid for small $\lambda$, and comes from 
taking the approximation for $D_{\text{err}}$ from eq. \req{D2smallla},
and neglecting $\langle \epsilon \delta \rangle$.
For example, for $\Delta \simeq \sigma_\delta$, we have a $4\%$
relative decrease in the sensitivity, ie
${\Delta \calT_{CMB}\over{ \calT_{CMB}}} \simeq 0.04$ within
this approximation (see \S \ref{sec:simulations}).
This loss of sensitivity only affects the noise (not the signal)
and could be partially
(or mostly) the result of the ADC hardware requierements,
rather than the compression process itself. 

\subsection{On Board Compression}

Romeo \etal (1998) have presented
a general study of (correlated multi-Gaussian) noise compression 
by studying  Shannon entropies per componet $h$, and therefore the
optimal compression $c_{r,opt}$ in Eq.(2).
For a linearly discretized data with
$n_{\text{bits}}=\log_2{N_{max}}$ bits, 
the Shannon entropy $h$ in Eq.[\ref{cr}]
depends only on the ratio of the digital resolution 
$\Delta$  to some
effective rms deviation, $\sigma_e$: 
\beq
h  = \log_2(\sqrt{2\pi e} ~\sigma_e/\Delta)
\eeq
with $\sigma_e^2 \equiv (\det C)^{1/N}$, 
where $C$ is the covariance
matrix for the (multi-Gaussian random) field $x_i$, i.e.,
$C_{ij} \equiv \langle x_i x_j \rangle$. In the case of the
error differences $\delta$ of Eq.[\ref{delta2}],
we have that $\sigma_e=\sigma_\delta$ and therefore:
\beq
h  = \log_2(\sqrt{2\pi e} ~/~\lambda)
\eeq

For a  data set with
$n_{\text{bits}}=\log_2{N_{max}}$ bits the
optimal compression rate in Eq.[\ref{cr}]] is  given by:
\beq
c_r \simeq {{n_{\text{bits}}\over{
\log_2 \left({\sqrt{2\pi e} ~\sigma_e/{\Delta}}\right)}}},
\label{crdelta}
\eeq
Thus, if we take $\Delta \simeq \sigma_\delta$
the optimal compression
is simply: 

\beq
c_{r,max}=n_{\text{bits}}/\log_2(\sqrt{2\pi e}) \simeq 8.
\label{crmax}
\eeq
where we have used $n_{\text{bits}}=16$ as planned for the Planck LFI.
This very large compression rate can be obtained because there is a large
range of values $\simeq \Delta 2^{n_{\text{bits}}}$  which
 has a very small probability, and therefore can be easily compressed
(e.g. by Huffman or arithmetic coding).
As mentioned above, the loss of sensitivity due to digital
distortion (i.e. Eq.[\ref{senCBM}]) is, 
in this case, $4\%$ within this approximation.

Another nice feature of our scheme is that
higher (or lower) compressions can be achieved if we are willing 
to reduce (or increase) the final temperature sensitivity to digital
distortion. As mentioned before this 
could be related to the ADC specifications.

\section{SIMULATIONS}
\label{sec:simulations}

\begin{figure}
\begin{center}
\begin{tabular}{c}
\epsfxsize 3.7in
\epsfbox{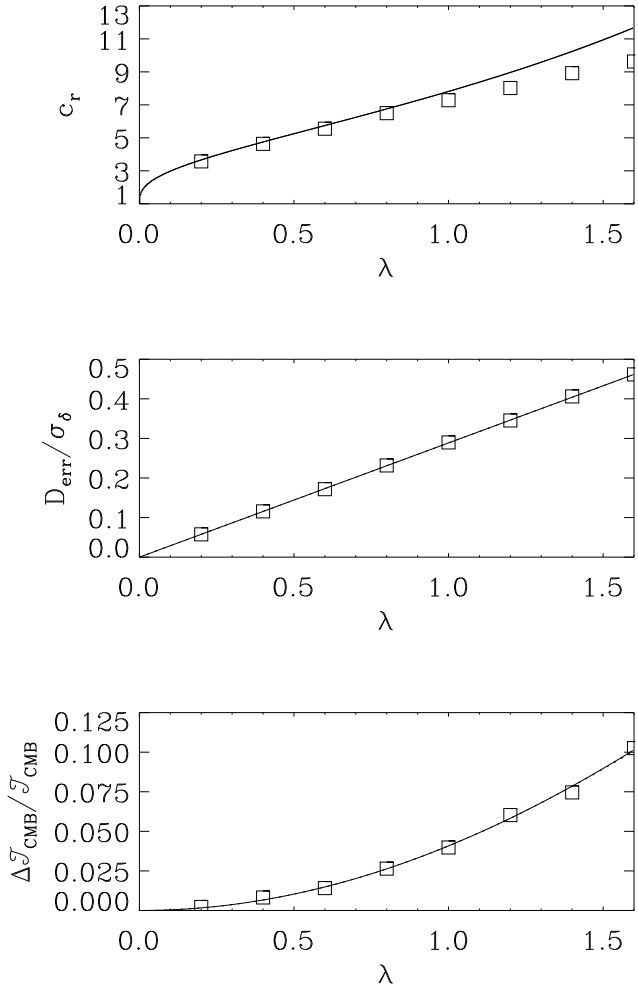}
\end{tabular}
\end{center}
\caption{
Discretization and compression simulations for a set
of $N=8700$ data values, quantized to different $\lambda$-values.
The three plots correspond to: compression factor $c_r$ (top), 
relative distorsion error ${ D_{\text{err}} \over \sigma_{\delta} }$ (middle),
and relative sensitivity variation
${ \Delta{\cal T}_{\text{CMB}} \over {\cal T}_{\text{CMB}} }
={ \sigma_{\hat{\delta}}-\sigma_{\delta} \over \sigma_{\delta} }$ 
(bottom).
Square symbols correspond to simulation results and
the solid lines have been obtained from the theoretical predictions
$c_r \simeq 16/h$, $ h \simeq \log_2\left( \sqrt{2 \pi \, e} / \lambda \right)$ (top),
$ { D_{\text{err}} \over \sigma_{\delta} } \simeq {\lambda \over \sqrt{12} }$
(middle) and
${\sigma_{\hat{\delta}}-\sigma_{\delta} \over \sigma_{\delta} } 
\simeq \sqrt{ 1+ {\lambda^2 \over 12} }-1$ (bottom).   
}
\label{FigSims}
\end{figure}

The process of generating, quantizing, storing, compressing, and
comparing the recovered and initial differences has been numerically
simulated.
A set of $\delta_{k,\alpha}$'s, $\alpha= 1,\dots,N$ for fixed
$k$, is produced as a random vector ---say $\delta$--- of Gaussian
components with a given variance $\sigma=\sigma_{\delta}$.
Next, the vector is linearly discretized or quantized, according to a
chosen value of $\lambda$, as explained in Romeo et al (1999),
yielding a new ---approximated--- vector called $\wh{\delta}$,
whose components are of the form
$\wh{\delta}_j= \delta_{\text{min}}+ q_j \Delta $, where $q_j$ is an integer. 
The set of values $q_j$, $j=1, \dots, N$, associated to each component, 
is then stored into a third vector made of 16-bit integers, and eventually written 
on a binary file, which is the object to be actually compressed.
A Huffmann compression program 
which has been specially adapted for 16-bit symbols is then applied to the file in question, 
and the resulting compression factor, 
which is the quotient between initial and final file sizes, is
duly recorded. The obtained compression factor is compared with the expected
theoretical result $\ds c_r={16 \over h}$, with $h$ given by eq.\req{crdelta}
with $\sigma_e=\sigma_{\delta}$.

Since Huffman compression involves no loss, the decoding procedure
amounts to recovering the $\wh{\delta}$ vector.
Therefore, the associated digital 
distortion error $D_{\text{err}}$ is nothing but 
the average of the
squared differences between the components of $\wh{\delta}$ and those
of $\delta$, as stipulated by eq. \req{defD2}. This distortion is
numerically evaluated, and its value compared with the small-$\lambda$
approximation given by eq. \req{D2smallla}.
Further, from the simualtions themsleves we find 
$\sigma_{\wh{\delta}}$, and calculate the sensitivity variation
as defined in eq. \req{senCBM}. The exact figures are compared
with the approximated part of the same equation. 

This is illustrated by the example depicted in Fig. \ref{FigSims}, 
which displays a simulation with
a Gaussian white noise vector of $N=8700$ components,
which corresponds to 1 minute of data at 6.9 ms sampling rate
(i.e. one circle). Up to
$\lambda \sim 1.2-1.5$, the actual compression factors are just
marginally smaller than the theoretical or ideal ones.
On the other hand, one can observe that
the small-$\lambda$ predictions (solid lines) for distortion and sensitivity
changes happen to be quite accurate.
The example shows that $c_r \sim 7.3$ for
$\lambda= 1 $, with a relative sensitivity decrease of
${\Delta {\cal T}_{\text{CMB}} \over {\cal T}_{\text{CMB}} }\sim 0.04$.

It is remarkable that the crude small-$\lambda$ approximations that have
been applied work so well for this problem.
To understand what happens, 
we have calculated corrections to these predictions 
by including:
\bit 
\item finite-sampling effects
\item the contribution of $\langle \epsilon \delta \rangle$ to 
$\sigma_{\wh{\delta}}$ 
\eit
Since we are handling finite samples, the integrations 
or summations of functions involving 
the probability distribution should be limited to the range effectively spanned 
by the available values of our stochastic variable. Given that we only have $N$ 
samples and a resolution limited by the value of $\lambda$, 
any magnitude of the order of ${1 \over N}$ will be indistinguishable from zero.
Hence, the {\it actual} range is just
$[-n(\lambda)\Delta, n(\lambda)\Delta]$,
where $n(\lambda)$ is determined by 
\beq 
f( n(\lambda)\Delta ) = {1 \over N} 
\eeq
and $f$ is our Gaussian probability distribution function.
This equality leads to
\beq 
n(\lambda)= \mbox{Round}\left[
\sqrt{ {2 \over\lambda^2} \ln\left( N\lambda \over \sqrt{2\pi} \right) }
\right] . 
\eeq
The $\langle \epsilon \delta \rangle$ correlation is so small that,	
up to now, it has been regarded as a vanishing quantity. 
However, if we take into account its nonzero value,    
the sensitivity variation will have to be evaluated according to the first line
of eq.\req{senCBM}. 
Both $D_{\text{err}}^2$ and $\langle \epsilon \delta \rangle$ have been calculated 
as sums of integrations between consecutive $\wh{\delta_n}$'s. 
Nevertheless, these sums are
not infinite, as $n$ ranges from $n= -n(\lambda)$ to $n= n(\lambda)$. 
The integration over each individual interval gives 	 
differences in incomplete gamma functions which have been numerically evaluated.
Not surprisingly, the result of applying all these corrections is very small indeed
(at $\lambda \sim 1$, they are of the order of $10^{-5}-10^{-4}$).
The corrected curves have been drawn in Fig. \ref{FigSims}
as dashed lines, but
they just overlap the existing lines
and can be hardly distinguished.

Another possibility is to perform a {\it nonlinear} quantization or
discretization.
Several tests have been made using a $\sinh(\alpha x)$ response function,
and changing the values of the nonlinearity parameter $\alpha$ 
(when $\alpha\to 0$,
the linear case is recovered). In general, the compression rate increases,
but the distortion becomes higher as well. For instance, when $\alpha=2.5$,
and the values of the discretization parameter are comparable 
to $\lambda \sim 1$, $c_r \sim 11$ and
$D_{\text{err}}/\sigma_{\delta} \sim 0.6$
(with linear discretization we had  $c_r \sim 7.3$ and
$D_{\text{err}}/\sigma_{\delta} \sim 0.3$). Taking $\alpha=5.0$, we find
$c_r \sim 13$ and $D_{\text{err}}/\sigma_{\delta} \sim 1.6$.
If we pick nonlinear and linear cases giving the same $c_r$, the distortion
associated to the linear one is, in general, smaller. Another disadvantage
of nonlinear quantization is that the mean of the discretized variable to be
stored may be too sensitive to the minimum and maximum values of $\eta$, which
can keep changing at every new set.  

\section{CONCLUSION}

We have considered several possible ways of reducing the size of
 the data on board
the Planck satellite:

\bit

\item (a) Averaging. One could average the information in
adjacent pixels within a circle or between consecutive images
of the same pixel.
\item (b) Changing the digital resolution, $\Delta$.
\item (c) Doing lossless compression.
\eit

Because of the existence of possible instrument instabilities 
and $1/f$-noise doing (a) alone, i.e., just averaging, could result 
in a dangerous decrease of the overall mission sensitivity. This
is illustrated in Eq.[\ref{sigma1f}] but will be better 
quantified in future studies.
Instead of this, one might try to use a low digital resolution, 
which should be balanced in order to maintain an acceptably low  digital 
distortion. 
A large digital distortion could bring about some loss of sensitivity, 
but this is more controlable than losses due to instrument instabilities
or lack of  diagnostic information.
The amount of possible lossless compression in (c) depends, in fact, 
on the digital resolution and on the statistical nature of the
signal (eg its Shannon entropy).
We have proposed to code the data in terms of
differences between  consecutive circles at a given sky pointing.
This technique allows for lossless compression and introduces
the flexibility to combine the above methods in a reliable way,
making precise predictions of how data can be compressed and
of how it could change the final mission sensitivity due to
digital distortion.
 
We have given some
quantitative estimates of how the above factors can be used to address the
problem of obtaining the large compression rates required for Planck.
For instance, one may observe the table below,
\begin{center}
\begin{tabular}{c|c|c}
$\lambda$&$c_r$&${\Delta {\cal T}_{\text{CMB}} \over {\cal T}_{\text{CMB}}}$ \\ \hline
0.6 & 5.6 & 0.01 \\
0.8 & 6.5 & 0.03 \\
1.0 & 7.3 & 0.04 \\
1.2 & 8.0 & 0.06 \\
1.4 & 8.9 & 0.07 \\
1.6 & 9.6 & 0.10 \\
\end{tabular}
\end{center}
taken from the simulation results shown in Fig. \ref{FigSims}. 
We have listed (Huffman) compression factors and sensitivity variations
for given values of the relative digital resolution parameter 
$\lambda=\Delta/\sigma$. At $\lambda=1$, a compression
rate of 7.3 has been found, at the price of increasing the theoretical
(continuous) sensitivity by $4\%$ due to the low digital resolution.
When $\lambda=1.6$, the compression 
reaches 9.6 and the sensitivity changes by just a 10$\%$. 

If we want to approach a
realistic modeling of the  final CMB map 
sensitivity we will need to know in detail which part of 
the diagnostic  on-board information should be downloaded to Earth.
More work is needed to find an optimal solution among the
different strategies listed above. 
The optimization will depend upon  other critical points of
the mission that still need to be specified in more
detail, such as: the survey and pointing strategy, the instrumental 
performance, the final temperature (or electric)
data ranges, the analogue-to-digital converters or the on board
calibration. We have argued that 
our proposal of coding and compressing the data in terms of
differences of consecutive circles at a given sky pointing,
has many advantages and is a first step towards this 
optimization.

\section*{Acknowledgments}

We would like to thank P.Fosalba, J.M. Herreros, 
R. Hoyland, R.Rebolo, R.A.Watson,
S.Levin and A. de Oliveira-Costa for discussions.
This work was in part supported by Comissionat per a Universitats i 
Recerca, Generalitat de Catalunya, grants ACES97-22/3 and ACES98-2/1,
and 1998BEAI400208 by the Spanish ``Plan Nacional del Espacio'', 
CICYT, grant ESP96-2798-E and by DGES (MEC, Spain), project PB96-0925.


                      
\appendix
\section{Appendix: RMS noise in a Gaussian difference}

\begin{table}
\begin{center}
\begin{tabular}{|c|c|c|c|}
\hline
\multicolumn{2}{|c|}{$f_{knee}=0.06$(Hz)} & \multicolumn{2}{|c|} 
{$f_{knee}=0.005$(Hz)}  \\
\hline
$f_{min}$ (Hz)& $\rho$ & $f_{min}$ (Hz) & $\rho$ \\
\hline
1/7200 & 9.8 $10^{-4}$ & 1/7200 & 8.1 $10^{-5}$  \\ \hline
3.17 $10^{-8}$ & 4.4 $10^{-3}$ & 3.17 $10^{-8}$ & 3.7 $10^{-4}$ \\ \hline 
\end{tabular}
\end{center}
\caption{Values of the autocorrelation $\rho$ between consecutive
sky pixels for different total calibration times, $1/f_{min}$.}
\label{table:rho}
\end{table} 

We can model the process of differencing as the subtraction of two 
gaussian random variables: $\eta_1$ and $\eta_2$ with variances
$\sigma_1^2$ and $\sigma_2^2$.
The probability density distribution for the difference 
random variable $\delta=\eta_2-\eta_1$ is also
a gaussian distribution with a new variance $\sigma_{\eta}^2$:
\begin{displaymath}
\sigma_{\delta}^2=\sigma_1^2 +\sigma_2^2-2\rho \sigma_1 \sigma_2.
\end{displaymath}
 For a wide sense stationary process $\sigma_1=\sigma_2=\sigma$ and
$\sigma_{\delta}^2= 2\sigma^2 ~(1-\rho)$.
One can obtain also in this way the expression for the entropy of the 
distribution,
\begin{displaymath}
h\approx \log_2 \left(\sqrt{2\pi 
e} ~\sigma_{\delta}/\Delta \right). 
\end{displaymath}
We want to take differences of data separated by $\tau = 1$ minute, which
corresponds to the same sky position. Bearing  in 
mind that our model is a 
first order Markov process $\rho$ will be equal to the correlation 
between 
pixels separated 1 min., that is $\rho \sigma^2 = C(\tau=1 ~\rm{min.})$ 
(recall that the 
correlation
matrix for a wide sense stationary stochastic process is a symmetric Toeplitz
matrix and so it depends only on index differences). Thus the
two-point correlation is:
\begin{displaymath} 
C(\tau)=
\int_{-\infty}^{+\infty}e^{i 2\pi f \tau }~P(f)~df.
\end{displaymath}
Next, we are going to estimate
this correlation for a power spectrum $P(f)$ of the type of 
white  noise plus $1/f$ (i.e. $P(f)$ in Eq.[\ref{pk}]).
In practice our spectrum will not run 
over the whole range but only  over a limited 
interval $(f_{min},f_{max})$. The final result,
for $\tau \neq 0$, is:
\begin{displaymath}
C(\tau)= 2~A~\left[
{\frac{\sin (2\pi f\tau)}{2\pi \tau}}
+f_{knee}~ ci(2\pi f\tau) \right]_{f=f_{min}}^{f=f_{max}} 
\end{displaymath}
where 
\begin{displaymath}
 ci(x)=-\int_{x}^{\infty}\frac{\cos t}{t} dt 
\end{displaymath}
We want to know if correlation due to $1/f$ noise is important in our data
handling model, so let us calculate some specific examples. In our model 
$f_{max}$ is given by the inverse of the sampling rate and $f_{min}$ is the 
inverse of two hours (if calibration occurs at every pointing) or the inverse
of the mission's time (about 1 year). 
In the Table \ref{table:rho} we have computed the magnitude of $\rho$ 
for our model and  for two different values of the calibration time, 
i.e. $1/f_{min}$. 
We can see in the Table how small the values of $\rho$
are compared to unity. Given the precision needed for the entropy
and compression factors, such contributions of $\rho$ to
$\sigma_\delta$ are neglegible.




\begin{thebibliography}{[00]}
\def\aj { ApJ, }
\def\aa {A \& A, }
\def\ajs{ ApJS, }
\def\mn { MNRAS, }
\def\apl { Ap. J. (Letters), }


\bibitem{Proposal}
Bersanelli, M. et al,
{\it COBRAS/SAMBA, Phase A Study for an ESA M3 Mission},
ESA report D/SCI(96)3.

\bibitem{Blum} Blum, E.J.
Annales d'Astrophysique, 22-2, 140, 1959

\bibitem{our2} Gazta\~naga, E., Barriga, J., Romeo, A.,
Fosalba, P., Elizalde, E. 1998, 
{\it Data compression on board the PLANCK Satellite Low Frequency Instrument: 
optimal compression rate}, Ap. Let. Com. in press,
astro-ph/9810205.

\bibitem{IAC} Herreros, J.M., Hoyland, R., Rebolo, R., Watson, R.A., 
1997 Ref.: LFI-IAC-TNT-001



\bibitem{Mandolesi} Mandolesi, N. \etal, 1998, LFI for Planck, 
a proposal to ESA's AO.


\bibitem{Maris} Maris, M.
{\it Planck LFI Consortium Meeting}, Florence, 1999 March 25-26

\bibitem{our1} Romeo, A.,  Gazta\~{n}aga, E., Barriga, J., 
Elizalde, E. 1998,
{\it Information content in Gaussian noise: optimal compression rates}, 
International Jounal of Modern Physics C, in press; physics/9809004.

\bibitem{GG}  Gersho A. and Gray,  R.M.,
{\it Vector Quantization and Signal Compression}, Kluwer Acad. Press, 1992.

\end{thebibliography}
\end{document}